# *ALICE*: THE *ROSETTA* ULTRAVIOLET IMAGING SPECTROGRAPH

S. A. Stern[1], D. C. Slater[2], J. Scherrer[2], J. Stone[2], M. Versteeg[2], M. F. A'Hearn[3], J. L. Bertaux[4], P. D. Feldman[5], M. C. Festou[*], Joel Wm. Parker[1], and O. H. W. Siegmund[6]

[1]*Southwest Research Institute, 1050 Walnut Street, Suite 426, Boulder, CO 80302-5143, USA*
[2]*Southwest Research Institute, 6220 Culebra Road, San Antonio, TX 78238-5166, USA*
[3]*University of Maryland, College Park, MD 20742, USA*
[4]*Service d'Aéronomie du CNRS, 91371 Verriéres le Buisson Cedex, France*
[5]*Johns Hopkins University, Department of Physics and Astronomy, Baltimore, MD 21218-2695, USA*
[6]*Sensor Sciences, 3333 Vincent Road, Pleasant Hill, CA 94523, USA*

**ABSTRACT**

We describe the design, performance and scientific objectives of the *NASA*-funded *ALICE* instrument aboard the ESA *Rosetta* asteroid flyby/comet rendezvous mission. *ALICE* is a lightweight, low-power, and low-cost imaging spectrograph optimized for cometary far-ultraviolet (FUV) spectroscopy. It will be the first UV spectrograph to study a comet at close range. It is designed to obtain spatially-resolved spectra of *Rosetta* mission targets in the 700-2050 Å spectral band with a spectral resolution between 8 Å and 12 Å for extended sources that fill its ~0.05° x 6.0° field-of-view.  *ALICE* employs an off-axis telescope feeding a 0.15-m normal incidence Rowland circle spectrograph with a concave holographic reflection grating. The imaging microchannel plate detector utilizes dual solar-blind opaque photocathodes (KBr and CsI) and employs a 2-D delay-line readout array. The instrument is controlled by an internal microprocessor. During the prime *Rosetta* mission, *ALICE* will characterize comet 67P/Churyumov-Gerasimenko's coma, its nucleus, and the nucleus/coma coupling; during cruise to the comet, *ALICE* will make observations of the mission's two asteroid flyby targets and of Mars, its moons, and of Earth's moon. *ALICE* has already successfully completed the in-flight commissioning phase and is operating normally in flight. It has been characterized in flight with stellar flux calibrations, observations of the Moon during the first Earth fly-by, and observations of comet Linear T7 in 2004 and comet 9P/Tempel 1 during the 2005 *Deep Impact* comet-collision observing campaign.

**1.0 INTRODUCTION**

Ultraviolet spectroscopy is a powerful tool for investigating the physical and chemical environments of astrophysical objects and has been applied with great success to the study of comets (e.g., Feldman 1982, Festou *et al.* 1993, Stern 1999, Feldman *et al.* 2004a; Bockelée-Morvan *et al.* 2004).  The *ALICE* UV spectrograph, designed to perform spectroscopic investigations of planetary atmospheres and surfaces at extreme (EUV) and far-ultraviolet (FUV) wavelengths between 700 and 2050 Å, has been optimized for *Rosetta* cometary science with high sensitivity, large instantaneous field-of-view, and broad wavelength coverage.

For the *Rosetta* Orbiter remote sensing investigation of comet 67P/Churyumov-Gerasimenko (67P/CG), *ALICE* will search for noble gases such as Ne and Ar; measure the production rates, variability, and structure of $H_2O$, CO, and $CO_2$ molecules that generate the bulk of cometary activity; measure the abundances and variability of the basic elemental species C, H, O, and N in the comet's coma; and measure atomic ion abundances in the comet's tail. In addition, *ALICE* will undertake an investigation of the FUV photometric properties of both the cometary nucleus itself, and solid grains entrained in the comet's coma. During *Rosetta*'s cruise to comet 67P/CG, *ALICE* will also obtain flyby observations of the mission's two target asteroids, the Earth's moon, and Mars. *ALICE* has already successfully completed a number of in-flight commissioning activities including in-flight calibration observations of several UV stars, and observations of comet C/2002 T7 (LINEAR) in April and May 2004 and comet 9P/Tempel 1 during the *Deep Impact* encounter in July 2005. By virtue of its location at the comet, the *ALICE* spectrograph will provide significant improvements in both sensitivity and spatial resolution over previous cometary UV observations.

---

[*] Deceased May 11, 2005

## 2.0 SCIENTIFIC OBJECTIVES AND CAPABILITIES

The scientific objectives of the *ALICE* investigation can be summarized as follows:

*(1) Search for and determine the evolved rare gas content of the nucleus to provide information on the temperature of formation and thermal history of the comet since its formation.* Some of the most fundamental cosmogonic questions about comets concern their place and mode of origin, and their thermal evolution since their formation. As comets are remnants from the era of outer planet formation, one of the most important things to be learned from them about planetary formation and the early solar nebula is their thermal history (see Bar-Nun *et al.* 1985; Mumma *et al.* 1993; Stern 1999). Owing to their low polarizabilities, the frosts of the noble gases are both chemically inert and extremely volatile. As a result, the trapping of noble gases is temperature dependent, so noble gases serve as sensitive "thermometers" of cometary thermal history. *ALICE* will determine (or set stringent limits) on the abundances of the He, Ne, Ar, Kr sequence from observations of their strongest resonance transitions at 584 Å (He I, to be observed in second order), 736/744 Å (Ne I), 1048/1067 Å (Ar I), and 1236 Å (Kr I). In addition to their importance as thermal history probes, evolved rare gas abundances also provide critical data for models requiring cometary inputs to the noble gas inventories of the planets. *ALICE* will also determine the abundances of another important low-temperature thermometer species, $N_2$ (electronic transitions in the 850-950 Å $c'_4$ and 1000-1100 Å Birge-Hopfield systems). He emission at 584 Å detected in comet C/1995 O1 (Hale-Bopp) by Krasnopolsky *et al.* (1997) from EUVE was accounted for by charge exchange of solar wind He ions. These authors also set a very stringent upper limit on Ne emission. Sensitive upper limits on Ar and $N_2$ have been obtained for four comets using FUSE (Weaver *et al.* 2002; Feldman 2005); a weak detection of Ar was published for Hale-Bopp (Stern *et al.* 2000). With long integration times *ALICE* should be able to detect and monitor several noble gases, and possibly also $N_2$, in comet 67P/CG without the *m/e* ambiguities of mass spectroscopy and at levels significantly below their cosmogonic abundance levels.

*(2) Determine the production rates of the parent molecule species, $H_2O$, CO and $CO_2$, and their spatial distributions near the nucleus, thereby allowing the nucleus/coma coupling to be directly observed and measured on many timescales.* Cometary activity and coma composition is largely driven by the sublimation of three key species: $H_2O$, CO, and $CO_2$. *ALICE* has been designed to directly detect each of these key parent molecules. Sunlight scattered by the nucleus, and probably background interplanetary H I Lyman-α, will be absorbed at a detectable level by water, and possibly by other molecules such as $CO_2$. By moving the slit to various locations around the coma (either by spacecraft pointing or simple changes in spacecraft location), it will be possible to map, even tomographically, the $H_2O$ distribution around the comet. By measuring the $H_2O$ column abundance in absorption in the UV, rather than by fluorescence (as in the IR), *ALICE's* measurements will provide a more direct, less model dependent signature for interpretation. CO will be observed via fluorescence in the well-known Fourth Positive band system from 1300 to 1700 Å, which will give total CO content; the CO Cameron bands between 1900 and 2050 Å will measure the CO produced by $CO_2$ photodissociation, and therefore the total $CO_2$ content. With most *ALICE* observations being directed toward the nucleus, the $H_2O$, CO, and $CO_2$ gas abundances will directly measure the production source(s) on the nucleus. The combination of these various types of observations will allow *ALICE* to completely map the $H_2O$ and CO distribution in the near environment of the nucleus and address the question of the coupling between the gas and the nucleus near the surface, and thus characterize the outgassing pattern of the surface. The latter will be particularly important for determining how much $H_2O$ and CO is derived from discrete, active regions, and how much from the "background" subsurface flow on the nucleus as a whole.

Any volatile species released on the night side of the nucleus will also be studied. When coupled with dust measurements in the coma, these data will yield information on the temporal and spatial variation of the dust/gas ratio in the cometary coma. When coupled with IR mapping measurements by *Rosetta*-VIRTIS, these data will yield information on the depths of the various icy reservoirs from which $H_2O$, CO, and $CO_2$ can be derived.

*(3)  Study the atomic budget of C, H, O, N, and S in the coma as a function of time.* As a UV instrument, *ALICE* is unique among the remote sensing investigations aboard *Rosetta* in its ability to detect atoms in the cometary atmosphere. Among the most important atomic species in comets are C, H, O, and N (Feldman *et al.* 2004a). *ALICE's* bandpass includes the strongest resonance lines of all of these (C I 1561, 1657, 1931 Å; H I 973, 1025, 1216 Å; O I 989, 1304 Å; and N I 1134 and 1200 Å), as well as those of another important, cosmogonically abundant species, S I (1425, 1474, 1813 Å). When *ALICE* is viewing the coma, we will obtain species abundance ratios. When the field-of-view encompasses the entire coma (e.g., during approach), the total content of the coma can be measured and its variation with time can be monitored. These quantities can be measured independently of any model. In addition, the most abundant coma ion, $O^+$, has its strongest resonance line at 834 Å that will allow an investigation of the ionization mechanism in the coma and the interaction of the comet ionosphere with the solar wind. Another observable ion, $C^+$ (1335 Å), will give complementary information on the competing ionization processes (photoionization vs. electron impact) as the two species are created from neutral atoms that are not similarly spatially distributed and the ions are thus differently affected by solar wind particles. And O I 1356 Å emission, though weak, will be an excellent tracer of electron impact processes in the coma. Together, these various probes will provide measurements of the abundances and spatial structures (e.g., distributed sources from CHON particles and as yet undetected organic molecules) in the atomic coma of comet 67P/CG. As a remote sensing instrument, *ALICE* enjoys the advantage that it can obtain such measurements throughout *Rosetta*'s long comet rendezvous, independently of the orbital location of the spacecraft.

*(4)  Study the onset of nuclear activity in ways* Rosetta *otherwise cannot.* ALICE is particularly well-suited to the exploration of one of the most fascinating cometary phenomena: the onset of nuclear activity. This area of interest has important implications for understanding cometary phenomenology, and for the general study of cometary activity at large distances (e.g., in Halley, Hale-Bopp, Skiff, Schwassmann-Wachmann 1, Chiron, etc.). *ALICE* will accomplish this by searching for and then monitoring the "turn-ons" of successively "harder volatiles" including $N_2$, $CO_2$, and $H_2O$ (in absorption), and the noble gases, CO, and atomic sulfur (all in fluorescence) as a tracer of $H_2S$ and $CS_2$.

*(5)  Spectral mapping of the entire nucleus of 67P/CG at FUV wavelengths in order to both characterize the distribution of UV absorbers on the surface, and to map the FUV photometric properties of the nucleus.* With its inherent long-slit imaging capability, *ALICE* can obtain either multispectral or monochromatic images of the comet at a resolution of $500/R_{50}$ meters, where $R_{50}$ is the comet-Orbiter range in units of 50 km. The UV images can be used to: (i) search for regions of clean ice; (ii) study the photometric properties of small ($10^{-9}$-$10^{-11}$ g) surface grains (both at and away from active zones) as a function of solar phase angle; and (iii) search for regions of electrical or photoluminescent glows on the surface. Further, by correlating regions that are dark below the ice absorption edge of $H_2O$ (1600-1700 Å), with visible albedo measurements made by the *Rosetta* imager OSIRIS, *ALICE* will be able to search for nuclear regions rich in this important volatile.

*(6)  Study the photometric and spectrophotometric properties of small grains in the coma as an aid to understanding their size distribution and how they vary in time.* UV photometry can be carried out with *ALICE* (a) using the solar continuum near 2000 Å, and (b) at H I Lyman-α (1216 Å). In both of these passbands, the photometric phase function of coma grains can be measured in order to map the distribution of grains with 10 to 100 times less mass than can be well-observed with the *Rosetta* imager, OSIRIS. We will separate the total optical depth so derived into icy and non-refractory components using the depth of the characteristic $H_2O$ absorption near 1650 Å as a compositional constraint.

*(7)  Map the spatial and temporal variability of $O^+$, $N^+$, $S^+$ and $C^+$ emissions in the coma and ion tail in order to connect nuclear activity to changes in tail morphology and structure near perihelion.* These ions have resonance transitions at 1036 Å and 1335 Å ($C^+$), 1085 Å ($N^+$), 910 Å and 1256 Å ($S^+$), and 834 Å ($O^+$), with which *ALICE* will be able to probe the ion formation and tail region behavior of the comet at any time when the comet is active.

## 3.0 TECHNICAL DESCRIPTION

### 3.1 Instrument Overview

An opto-mechanical layout of *ALICE* is shown in Figure 1. Light enters the telescope section through a 40 x 40 mm$^2$ entrance aperture and is collected and focused by an f/3 off-axis paraboloidal (OAP) mirror onto the entrance slit and then onto a toroidal holographic grating, where it is dispersed onto a microchannel plate (MCP) detector that uses a double-delay line (DDL) readout scheme. The 2-D (1024 x 32)-pixel format, MCP detector uses dual, side-by-side, solar-blind photocathodes: potassium bromide (KBr) and cesium iodide (CsI). The measured spectral resolving power ($\lambda/\Delta\lambda$) of *ALICE* is in the range of 70-170 for an extended source that fills the instantaneous field-of-view (IFOV) defined by the size of the entrance slit. *ALICE* is controlled by an SA 3865 microprocessor, and utilizes lightweight, compact, surface mount electronics to support the science detector, as well as the instrument support and interface electronics. Figure 2 shows both a 3D external view of *ALICE*, and a photograph of the flight unit.

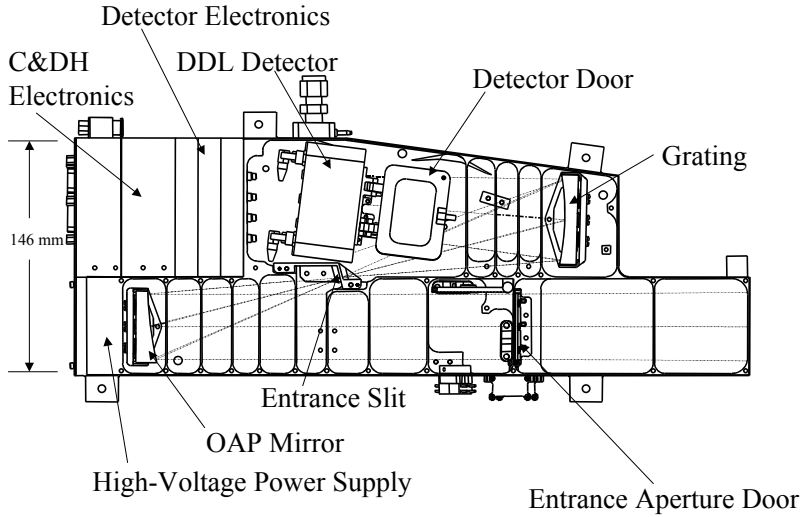

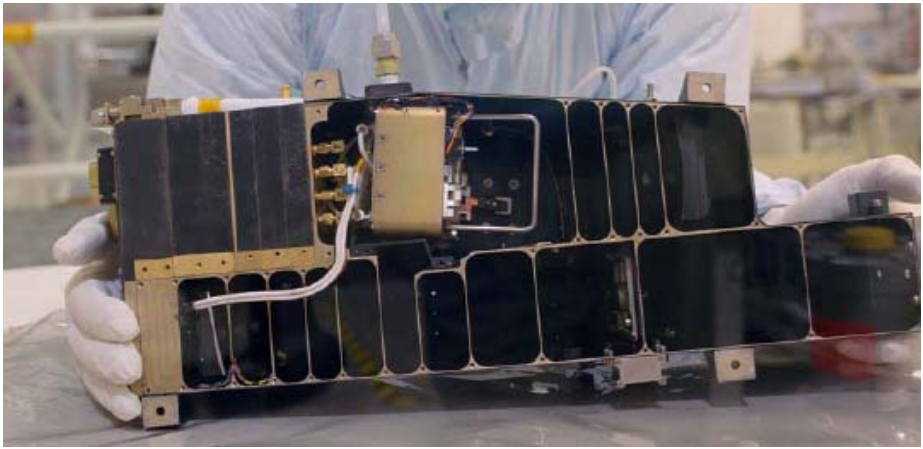

*Fig. 1.* (a) The opto-mechanical layout of ALICE. (b) A photograph of the ALICE flight unit.

### 3.2 Optical Design

The OAP mirror has a clear aperture of 41 x 65 mm$^2$, and is housed in the telescope section of the instrument (see Figures 1 and 2). The reflected light from the OAP enters the spectrograph section, which contains a holographic grating and MCP detector. The slit, grating, and detector are all arranged on a 0.15-m diameter normal incidence Rowland circle.

The spectrograph utilizes the first diffraction order throughout the 700-2050 Å spectral passband. The lower half of the first order wavelength coverage (700-1025 Å) also shows up in second order between the first order wavelengths of 1400 and 2050 Å.

Both the OAP and grating, and their mounting fixtures, are constructed from monolithic pieces of Al, coated with electroless Ni and polished using low-scatter polishing techniques. The OAP and grating optical surfaces are overcoated with sputtered SiC. Control of internal stray light is achieved with a well-baffled optical cavity, and a holographic diffraction grating that has low scatter and near-zero line ghost problems.

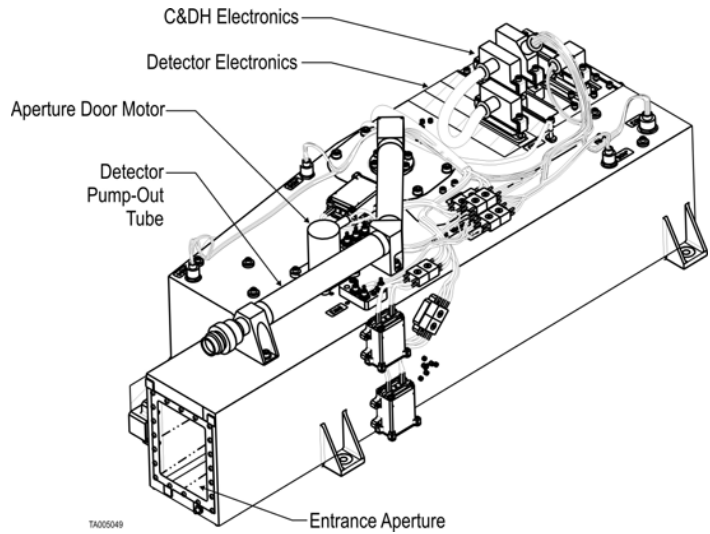

(a)

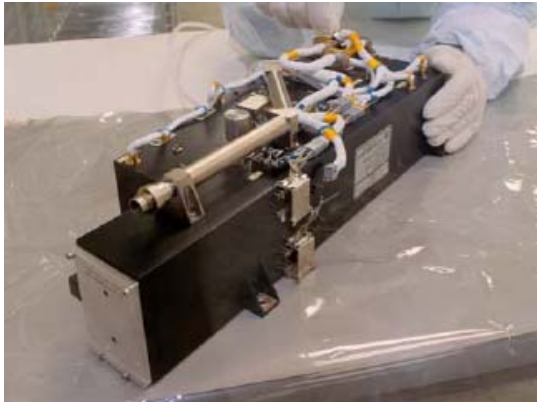

(b)

*Fig. 2. (a) 3D external view of* ALICE. *(b) Photograph of the* ALICE *flight unit.*

For contamination control, heaters are mounted to the back surfaces of the OAP mirror and grating to prevent cold trapping of contaminants during flight. To protect the sensitive photocathodes and MCP surfaces from exposure to moisture and other harmful contaminants during ground operations, instrument integration, and the early stages of the mission, the detector tube body assembly is enclosed in a vacuum chamber with a front door that was successfully (and permanently) opened during the early commissioning phases of the flight. For additional protection of the optics and detector from particulate contamination during the flight, a front entrance aperture door is included that can close when the dust and gas levels are too high for safe operation and exposure (i.e., when the *Rosetta* Orbiter is close to the comet nucleus). The telescope baffle vanes also help to shield the OAP mirror from bombardment of small particles that can enter the telescope entrance aperture.

### 3.3 Entrance Slit Design

The spectrograph entrance slit assembly design is shown in Figure 3. The slit is composed of three sections plus a pinhole mask. The center section of the slit provides high spectral resolution of ~8-12 Å FWHM with an IFOV of 0.05° x 2.0°. Surrounding the center slit section are the two outer sections with IFOVs of 0.10° x 2.0° and 0.10° x 1.53°. A pinhole mask, located at the edge of the IFOV of the second outer section, provides limited light throughput to the spectrograph for bright point source targets (such as hot UV stars) that will be used during stellar occultation studies of CG's coma.

### 3.4 Detector & Detector Electronics

The 2-D imaging photon-counting detector located in the spectrograph section utilizes an MCP Z-stack that feeds the DDL readout array (Siegmund *et al.* 1992). The input surface of the Z-stack is coated with opaque photocathodes of KBr (700-1200 Å) and CsI (1230-2050 Å) (Siegmund *et al.* 1987). The detector tube body is a custom design made of a lightweight brazed alumina-Kovar structure that is welded to a housing that supports the DDL anode array (see Figures 4 and 5).

To capture the entire 700-2050 Å bandpass and 6° spatial FOV, the size of the detector's active area is 35 mm (in the dispersion direction) x 20 mm (in the spatial dimension), with a pixel format of (1024 x 32)-pixels. The 6° slit-height is imaged onto the central 20 of the detector's 32 spatial channels; the remaining spectral channels are used for dark count monitoring. Our pixel format allows Nyquist sampling with a spectral resolution of ~3.4 Å, and a spatial pixel resolution of 0.3°.

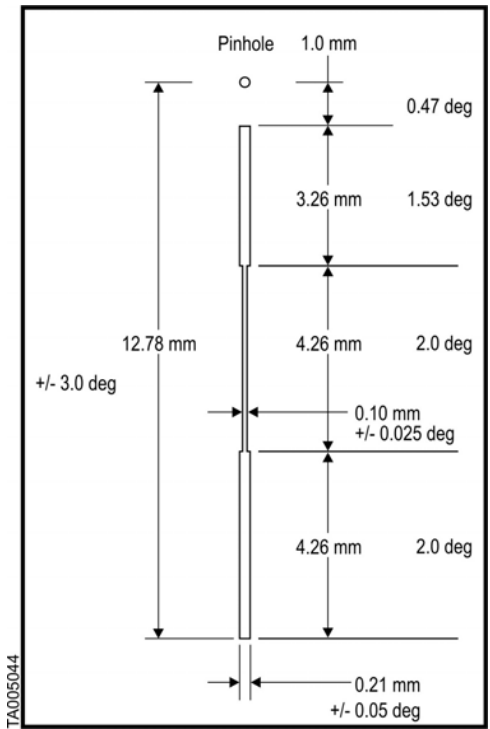

*Fig. 3.* ALICE *entrance slit design.*

The MCP Z-Stack is composed of three 80:1 length-to-diameter (L/D) MCPs that are all cylindrically curved with a radius-of-curvature of 75 mm to match the Rowland-circle for optimum focus across the full spectral passband. The total Z-Stack resistance at room temperature is ~500 MΩ. The MCPs are rectangular in format (46 x 30 mm$^2$), with 12-μm diameter pores on 15-μm centers. Above the MCP Z-Stack is a repeller grid that is biased ~1000 volts more negative than the top of the MCP Z-Stack. This repeller grid reflects electrons liberated in the interstitial regions of the MCP back down to the MCP input surface to enhance the detective quantum efficiency of the detector.

The expected H I Lyman-α (1216 Å) emission brightness from comet 67P/CG is ~4 kR at a heliocentric distance of 1.3 AU (based on *IUE* observations of this comet in 1982; Feldman *et al.* 2004b). To prevent saturation of the detector electronics, it is necessary to attenuate the Lyman-α emission brightness to an acceptable count rate level well below the maximum count rate capability of the electronics (i.e., below $10^4$ c s$^{-1}$). An attenuation factor of at least an order of magnitude is required to achieve this lower count rate. This was easily achieved by physically masking the MCP active area where the H I Lyman-α emission comes to a focus during the photocathode deposition process. The bare MCP glass has a quantum efficiency about 10 times less than that of KBr at 1216 Å.

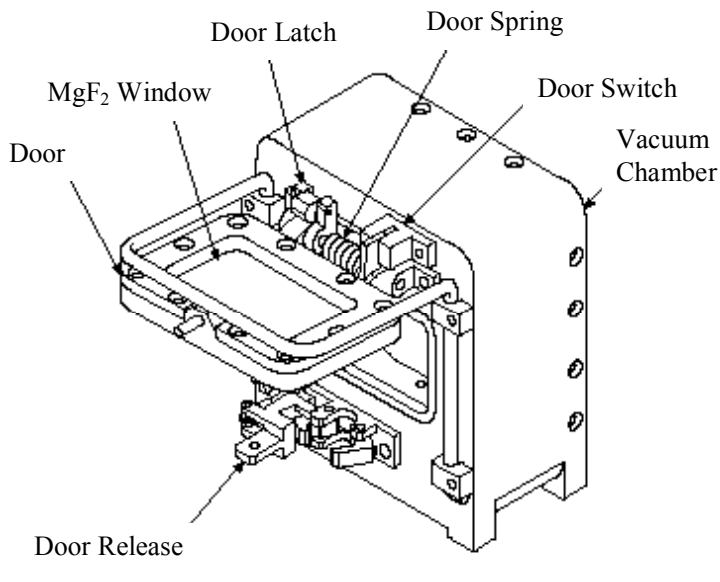

*Fig. 4.* Schematic of the ALICE *DDL detector vacuum chamber housing.*

Surrounding the detector tube body is the vacuum chamber housing made of aluminum and stainless steel (see Figures 4 and 5). As mentioned above, this vacuum chamber protected the KBr and CsI photocathodes against damage from moisture exposure during ground handling and from outgassing constituents during the early stages of the flight. It also allowed the detector to remain under vacuum (< $10^{-5}$ Torr) during ground operations, testing and handling, and transportation. Light enters the detector vacuum chamber through an openable door, which contains a built-in MgF$_2$ window port that transmits UV light at wavelengths > 1200 Å. This window allowed testing of the detector with the door closed, and provided redundancy during flight if the door mechanism had failed to open.

The detector vacuum chamber door was designed to be opened once during flight, using a torsion spring released by a dual-redundant pyrotechnic actuator (dimple motor). During instrument integration and test (I&T), the door was successfully opened numerous times and manually reset. In flight, the detector was successfully opened; however, the primary side of the actuator did not open the door—the redundant side was required to successfully open the door.

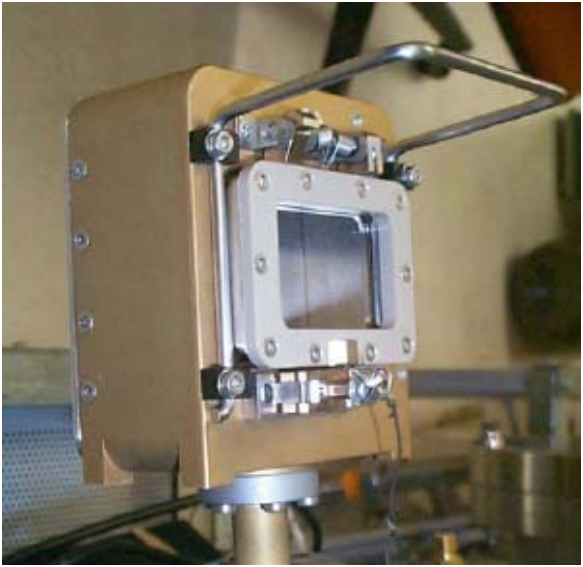

*Fig. 5. A photograph of the* ALICE *DDL flight detector with the MgF$_2$ detector door in the closed position.*

The detector electronics includes preamplifier circuitry, time to digital converter circuitry (TDC), and pulse-pair analyzer (charge analysis) circuitry (PPA). All of these electronics are packaged into three 64 x 76 mm$^2$ boards. These three boards are mounted inside a separate enclosed magnesium housing that mounts to the rear of the spectrograph section (just behind the detector vacuum chamber). The detector electronics require ± 5 VDC, and draw ~1.1 W.

The detector electronics amplify and convert the detected output pulses from the MCP Z-Stack to pixel address locations. Only those analog pulses output from the MCP that have amplitudes above a set threshold level are processed and converted to pixel address locations. For each detected and processed event, a 10-bit *x* address and a 5-bit *y* address are generated by the detector electronics and sent to the *ALICE* command-and-data handling (C&DH) electronics for data storage and manipulation. In addition to the pixel address words, the detector electronics also digitizes the analog amplitude of each detected event output by the preamplifiers and sends this data to the C&DH electronics. Histogramming this "pulse-height" data creates a pulse-height distribution function that is used to monitor the health and status of the detector during operation. A built-in "stim-pulser" is also included in the electronics that simulates photon events in two pixel locations on the array. This pulser can be turned on and off by command and allows testing of the entire *ALICE* detector and C&DH electronic signal path without having to power on the detector high-voltage power supply. In addition, the position of the stim pixels provides a wavelength fiducial that can shift with operational temperature.

### 3.5 Electrical Design

The instrument support electronics (see Figure 6) on *ALICE* include the power controller electronics (PCE), the C&DH electronics, the telemetry/command interface electronics, the decontamination heater system, and the detector high-voltage power supply (HVPS). All of these systems are controlled by a rad-hardened SA 3865 microprocessor, supplied by Sandia Associates, with 32 KB of local program RAM and 64 KB of acquisition RAM along with 32 KB of program ROM and 128 KB of EEPROM. All of the instrument support electronics are contained on 5 boards mounted just behind the detector electronics (see Figures 1 and 2).

*Power Controller Electronics.* The PCE are composed of DC/DC converters designed to convert the spacecraft power to ± 5 VDC required by the detector electronics, the C&DH and TM interface electronics, and the detector HVPS. Also located in the PCE is the switching circuit for the heaters and the limited angle torque (LAT) motor controller that operates the front aperture door.

*Command-and-Data Handling Electronics.* The C&DH electronics handles the following instrument functions: (i) the interpretation and execution of commands to the instrument, (ii) detector acquisition control including the histogramming of raw detector event data, (iii) telemetry formatting of both science and housekeeping data, (iv) control of the detector HVPS, (v) the detector vacuum cover release mechanism, (vi) the front aperture door control, (vii) the control of the housekeeping ADC's which are used to convert analog housekeeping data to digital data for inclusion into the TM data stream, and (viii) on-board data handling.

*Telemetry/Command Interface Electronics.* The C&DH utilizes radiation tolerant buffers and FIFO memory elements in the construction of the spacecraft telemetry and command interfaces. A finite state machine programmed into a radiation hardened Actel 1280 FPGA controls the receipt and transmission of data. A bit-serial interface is used.

*Decontamination Heater System.* A single decontamination heater each (~1 W resistive heater) is bonded to the backside surface of both the OAP mirror substrate, and the grating substrate. Along with each heater, two redundant thermistors are also mounted to the back of each substrate to monitor and provide control feedback to the heaters. The C&DH electronics can separately control each heater. Successful heater activations have already taken place during the commissioning phase of the flight. Additional activations are planned periodically during the long cruise phase to comet 67P/CG.

*High Voltage Power Supply.* The HVPS is located in a separate enclosed bay behind the OAP mirror (see Figure 1). It provides the –4.0 kV required to operate the detector. The voltage to the Z-stack is fully programmable by command in ~25 V steps between –1.7 and –6.1 kV. The mass of the supply is ~120 g, and consumes a maximum of 0.65 W during detector operation.

### 3.6    Data Collection Modes

*ALICE* can be commanded to operate in one of three data collection modes: i) image histogram, ii) pixel list, and iii) count rate modes. Each of these modes uses the same 32k word (16 bit) acquisition memory. The first two acquisition modes use the same event data received form the detector electronics but the data is processed in a different way.  Also, in these two modes, events occurring in up to eight specific areas (each area is composed of 128 spectral pixels by 4 spatial pixels) on the array can be excluded to isolate high count rate areas that would otherwise fill up the array.  The third acquisition mode only uses the number of events received in a given period of time; no spectral or spatial information is used.

*Image Histogram Mode.*   In this mode, acquisition memory is used as a two dimensional array with a size corresponding to the spectral and spatial dimensions of the detector array.  The image histogram mode is the prime *ALICE* data collection mode (and the one most often used during flight).  During an acquisition, event data from the detector electronics representing ($x,y$)-pixel coordinates are passed to the histogram memory in parallel form.  The parallel data stream of $x$ and $y$ values is used as an address for a 16 bit cell in the 1024 x 32 element histogram memory, and a read-increment-write operation on the cell contents is performed for each event. During a given integration time, events are accumulated one at a time into their respective histogram array locations creating a 2-D image.  The read-increment-write operation saturates at the maximum count of 65,535 so no wrap around can occur in the acquired data.  The time information of the individual events within the acquisition is lost in this process, but using appropriate acquisition durations high signal-to-noise ratio data may be acquired even from dim objects.  At the conclusion of the integration period the acquired data can be down linked in telemetry.  In order to limit the required telemetry bandwidth, the histogram memory can be manipulated to extract only data from up to eight separate, 2-D windows in the array for downlink, and within these windows rows and columns may be co-added to further reduce the number of samples.

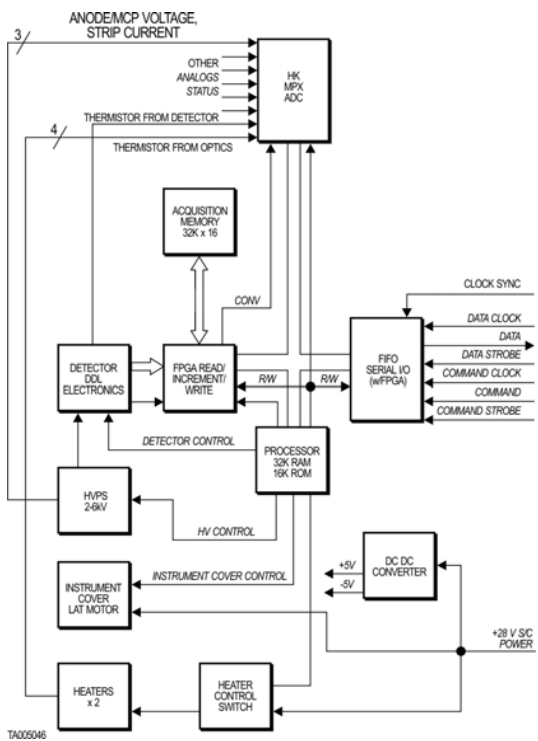

***Fig. 6.*** ALICE's *electronic block diagram.*

*Pixel List Mode.*  In this mode the acquisition memory is used as a one-dimensional linear array of 32,768 entries. The pixel list mode allows for the sequential collection of each ($x,y$)-event address into the linear pixel list memory array. Periodically, at programmable rates not exceeding 256 Hz, a time marker is inserted into the array to allow for "time-binning" of events. This mode can be used to either (a) lower the downlink bandwidth for data collection integrations with very low counting rates, or (b) for fast time-resolved acquisitions using relatively bright targets in the *ALICE* FOV. At the conclusion of this acquisition period the total amount of generated data can be further reduced by selecting only events that have occurred within up to eight separate windows for downlink.

*Count Rate Mode.* In this mode the acquisition memory is again used as a one dimensional linear array of 32,768 entries. The count rate mode is designed to periodically (configurable between 3 ms and 12 s) collect the total detector array count rate sequentially in the linear memory array, as if the entire instrument were an FUV photometer. This mode allows for high count rates from the detector (up to 10 kHz), without rapid fill up of the array. It does not, however, retain any spatial or spectral information for broadband photometric studies. Depending on the required periodic acquisition rate, total acquisition durations of up to 98 seconds to 100 hours are possible.

## 4.0 MEASURED PERFORMANCE CHARACTERISTICS

An overview of the instrument characteristics, spacecraft resource requirements, and a comparison of the ground and in-flight performance of *ALICE* are summarized in Table I. A brief summary of the ground test and in-flight radiometric performance is presented below.

### 4.1 Radiometric Ground Test

*ALICE*'s radiometric performance was first measured prior to delivery to the spacecraft. Both vacuum and bench level radiometric tests allowed characterization of the detector dark count rate, the instrument's wavelength passband, spatial and spectral resolution, scattered light rejection, and the effective area as a function of wavelength (see Slater *et al.* 2001 for details of these test results). The vacuum tests were performed using the UV radiometric test facility at Southwest Research Institute (SwRI) following the successful completion of instrument environmental tests that included vibration, EMI, and thermal-vacuum tests. Table I includes a summary of the *ALICE* radiometric performance measured during these ground tests, with a comparison to the in-flight performance measured during the commissioning phase of the flight. Note that the ground and in-flight performance tracks closely except for i) the detector background rates, which are higher in flight as expected due to the ambient spacecraft environment; and ii) the effective area, which was found to be lower in flight by a factor of two from that measured during ground test. The lower in-flight effective area performance is not well understood; speculation includes possible degradation of the detector photocathodes during the extended storage of the spacecraft caused by the one-year launch delay.

### 4.2 In-Flight Performance

During the commissioning phase of flight, *ALICE* performed a number of in-flight radiometric characterization and calibration observations using UV stars (focus, PSF, effective area, pointing, wavelength calibration), the Sun (stray light characterization), and the Moon during the first Earth-Moon flyby (effective area). Table I summarizes the in-flight instrument performance characteristics. Figure 7 shows the measured in-flight effective area based on a number of UV stars observed with *ALICE* that have been calibrated with IUE. Effective area values using *ALICE* acquired lunar spectral data also fits well with the data acquired using UV-bright stars shown in Figure 7.

## 5.0 CONCLUSION

*ALICE* is a highly-capable, low-cost UV imaging spectrograph that will significantly enhance *Rosetta's* scientific characterization of the nature and origins of the cometary nucleus, its coma, and nucleus/coma coupling. *ALICE* will do this by its study of noble gases, atomic abundances in the coma, major ion abundances in the tail, and powerful, unambiguous probes of the production rates, variability, and structure of $H_2O$ and $CO/CO_2$ molecules that generate cometary activity, and the far-UV properties of the nucleus and solid grains. *ALICE* will also deepen the *Rosetta* Orbiter's in situ observations by giving them the global view that only a remote-sensing adjunct can provide.

The in-flight characterization and calibration performance of *ALICE* has successfully been completed and the instrument is healthy and ready to accomplish scientific tasks aboard the *Rosetta* Orbiter.

**TABLE I**
*Rosetta-ALICE* **Characteristics, Spacecraft Resource Requirements, & Measured Performance Summary (Ground cal & in-flight results).**

| Parameter | Description |
|---|---|
| Total Spectral Passband: | 680 – 2060 Å |
| Spectral Resolution: | (Point Source) Grd Cal: 4-8 Å; Flight: 4-9 Å |
| | (Extended Source) Grd Cal and Flight: 8-12 Å |
| Spatial Resolution: | 0.05° x 0.6° (35 x 420 m$^2$ at 40 km from nucleus) |
| Active FOV | 0.05° x 2.0° + 0.1° x 2.0° + 0.1° x 1.5° |
| Pointing | Boresight with OSIRIS WAC and VIRTIS |
| Effective Area: | Flight: 0.02 (1575 Å)-0.05 cm$^2$ (1125 Å) |
| Stray Light Attenuation | Grd Cal: $< 10^{-4}$ at $\theta_{off} > 4°$; Flight: $< 10^{-9}$ at $\theta_{off} > 60°$ |
| Detector Dark Rate: | Grd Cal: 2.4 c/s; Flight: 18 c/s (total array) |
| Telescope/Spectrograph | Off-axis telescope, Rowland circle spectrograph |
| Detector Type | 2-D Microchannel Plate w/ double-delay line readout |
| External Dimensions | 204 x 413 x 122 mm$^3$ |
| Mass/Power | 3.0 kg/4.0 W |
| Observation Types | Nucleus imaging and spectroscopy; coma gas spectroscopy; jet and grain spectrophotometry; stellar occultations (optional observations) |

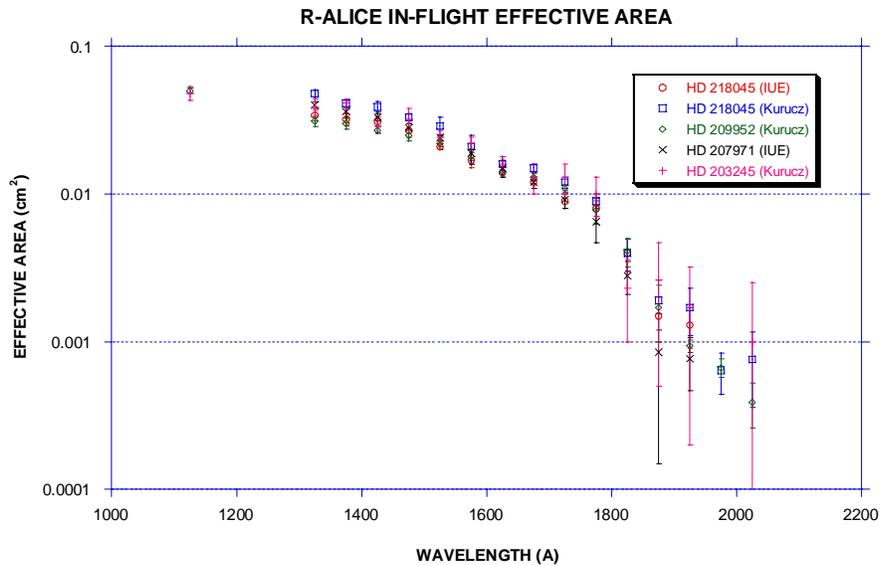

*Fig. 7. The measured in-flight effective area of* ALICE *based on observations of various UV stars during the* Rosetta *instrument commissioning phase of the mission.*

## Acknowledgements

We would like to thank our engineering staff at SwRI, including Greg Dirks, Susan Pope, and Peter De Los Santos, for their contributions to the design of *ALICE*. We also want to thank Dr. John Vallerga and Rick Raffanti of Sensor Sciences for their technical advice and assistance with the detector and its associated electronics design and test, and Joe Kroesche for the initial design of the *ALICE* flight software.